\documentclass[prb,twocolumn,showpacs,preprintnumbers,amsmath,amssymb]{revtex4}

\usepackage{graphicx}
\usepackage{dcolumn}
\usepackage{bm}
\usepackage{mathrsfs}
\usepackage{subfigure}
\usepackage{natbib}

\begin{document}
\title{Different length-scales for order parameters in two-gap superconductors: \\the extended Ginzburg-Landau theory}

\author{L. Komendov\'{a}}
\affiliation{Departement Fysica, Universiteit Antwerpen,
Groenenborgerlaan 171, B-2020 Antwerpen, Belgium}

\author{M. V. Milo\v{s}evi\'{c}}
\affiliation{Departement Fysica, Universiteit Antwerpen,
Groenenborgerlaan 171, B-2020 Antwerpen, Belgium}

\author{A. A. Shanenko}
\affiliation{Departement Fysica, Universiteit Antwerpen,
Groenenborgerlaan 171, B-2020 Antwerpen, Belgium}

\author{F. M. Peeters}
\affiliation{Departement Fysica, Universiteit Antwerpen,
Groenenborgerlaan 171, B-2020 Antwerpen, Belgium}
\date{\today}

\begin{abstract}
Using the Ginzburg-Landau theory extended to the next-to-leading
order we determine numerically the healing lengths of the two order
parameters at the two-gap superconductor/normal metal interface. We
demonstrate on several examples that those can be significantly
different even in the strict domain of applicability of the
Ginzburg-Landau theory. This justifies the use of this theory to
describe relevant physics of two-gap superconductors, distinguishing
them from their single-gap counterparts. The calculational degree of
complexity increases only slightly with respect to the conventional
Ginzburg-Landau expansion, thus the extended Ginzburg-Landau model
remains numerically far less demanding compared to the full
microscopic approaches.
\end{abstract}


\maketitle

Over the past half-century, the Ginzburg-Landau (GL)
theory~\cite{GL} has proven to be a very helpful tool in studies of
superconductors, but also other systems in and out of the condensed
matter physics. By its construction, the formalism is only justified
near the critical temperature $T_c$, but it typically produces
qualitatively correct results even far below the $T_c$. Recently an
exception was found to this unwritten rule, when it was shown that
the standard formulation of the GL theory is insufficient for
adequate description of two-band (or multi-band) superconductors
because it predicts the same spatial variation of the condensates in
all bands\cite{Kogan}(within its range of applicability i.e. close
to $T_c$). This renders it unable to connect to the theoretical
results obtained by using Usadel\cite{Koshelev2003, Koshelev2004} or
Eilenberger\cite{Silaev2011} equations in the broader temperature
range, which unambiguously show presence of two separate length
scales for two gaps. Experimentally, the evidence for different
coherence lengths in $\sigma$ and $\pi$ bands of MgB$_2$ was
previously found by the direct vortex imaging\cite{Eskildsen} and in
pronounced features in the behaviour of the thermal
conductivity\cite{Sologubenko}, specific heat\cite{Bouquet} and
flux-flow resistivity\cite{Shibata} as a function of applied
magnetic field.

In order to capture the important physics of different length
scales, one needs to extend the GL formalism as realized in
Ref.~\onlinecite{Arkady}, where the two order parameters are
calculated up to order $\tau^{3/2}$ in the small deviation from the
critical temperature $\tau=1-T/T_c$, instead of the standard
$\tau^{1/2}$ as used in Ref.~\onlinecite{Kogan}. In the latter
article, the authors mention that the extra terms in the
next-to-leading order are by construction small corrections, not
significant enough to alter the {\it single} coherence length
controlling the spatial distribution of {\it both} condensates.
First, we argue that the effects of higher-order corrections can be
significant since the above argument about small corrections applies
only to the order parameter - not necessarily to its spatial profile
i.e. healing lengths and other physical quantities. Second, we argue
that any (even small) difference between the characteristic length
scales of two condensates is of fundamental importance, since it may
lead to other novel phenomena related to the competition of length
scales. Note that here we do not enter the recent debate about
sufficient discrepancy of length scales to provide `type-1.5'
superconductivity \cite{comm}. Instead, we complement that
discussion, by exactly quantifying the difference in length scales
in the domain of the extended GL theory.

The fact that the difference of characteristic length scales of two
Cooper pair condensates exists and can be significant even in the
strict Ginzburg-Landau domain is of great practical importance,
since the calculations based on the Ginzburg-Landau theory are
typically far less computationally demanding than the calculations
based on full microscopic theories (Bogoliubov-de Gennes, Gor'kov,
Usadel or Eilenberger equations). In microscopic formalisms, one
usually has to make clever approximations and make compromises in
the calculation procedure. As a consequence, even though these
approaches are valid in the whole temperature range from absolute
zero to $T_c$, they are limited to the simple systems such as a
single vortex or other highly symmetric or effectively
one-dimensional cases. On the contrary, the Ginzburg-Landau
equations have a much simpler structure and therefore allow for
studying of highly non-trivial situations such as vortex lattice
statics and dynamics, current driven systems, interaction with
pinning and fluxonic devices. Since the above is well established in
the standard GL formalism, we emphasize here that the calculations
become only a fraction more complicated in the extended
Ginzburg-Landau formalism - while it does contain more equations
(for the order parameters to the leading, and next-to-leading
order), the coupling of those equations is realized through the
coefficients and not the calculated variables - contrary to the
standard set of two-band GL equations \cite{roel}.

\paragraph*{Theoretical formalism} Following Ref.~\onlinecite{Arkady}, we
employ the extended GL formalism for the order parameters $\Delta_j$
evaluated up to order $\tau^{3/2}$ by taking $\Delta_j(\mathbf x) =
\Delta_j^{(0)}(\mathbf x)+\Delta_j^{(1)}(\mathbf x)$, where
$\Delta_j^{(0)}(\mathbf x) \propto \tau^{1/2}$ and
$\Delta_j^{(1)}(\mathbf x) \propto \tau^{3/2}$, with $j=1,2$
indexing two coupled condensates in a two-band superconductor. In
the absence of applied magnetic field, the order parameters can be
taken real. The extended GL equations then read (at zero magnetic
field)
\begin{subequations}\label{gleqdim}
\begin{align}
&\alpha \Delta_j^{(0)} + \beta_j[\Delta_j^{(0)} ]^3 - K\nabla^2 \Delta_j^{(0)} = 0, \label{gleqdim_A}\\
&\Delta_j^{(1)}\big(\alpha +3\beta_j[\Delta_j^{(0)} ]^2 \big)- K\nabla^2 \Delta_j^{(1)}=\nonumber \\
&\quad\quad\quad\quad\quad\quad\quad\quad\quad F(\Delta_j^{(0)})+F_j(\Delta_j^{(0)}),
\label{gleqdim_B}
\end{align}
\end{subequations}
where $\alpha=- N(0) \tau\,\big(n_1 {\cal A}_2 + n_2 {\cal
A}_1\big)/\lambda_{12},~K=N(0) \hbar^2 W^2_3\big(n_1  v^2_1 {\cal
A}_2 + n_2  v^2_2{\cal A}_1\big)/(6\lambda_{12}),~\beta_{1(2)}=N(0)
W^2_3 \Big(n_{1(2)}{\cal A}_{2(1)}/\lambda_{12} +n_{2(1)}{\cal
A}^3_{1(2)}/\lambda^3_{12}\Big)$, with ${\cal A}_1 = \lambda_{22} -
\eta n_1 {\cal A}$ and ${\cal A}_2 = \lambda_{11} - \eta n_2 {\cal
A}$. Here ${\cal A}=\ln(2e^{\Gamma}\hbar\omega_D/\pi T_c)$, with
Euler constant $\Gamma$= 0.577, and $\eta$ denotes the determinant
of the interaction matrix $\lambda_{ij} = N(0)g_{ij}$, with $g_{ij}$
the coupling constant, $N(0)$ the total density of states (DOS) and
$n_jN(0)$ the band-dependent DOS. In addition, the coefficients
feature Fermi velocities $v_j$ of both bands,
$W^2_3=\frac{7\zeta(3)}{8\pi^2T_c^2}$, with $\zeta(\ldots)$ the
Riemann zeta-function, and terms $F$ and $F_j$ which are given in
complete form in Ref.~\onlinecite{Arkady}. Since
Eq.~(\ref{gleqdim}a) is completely equivalent to the single-gap GL
equation and the bands share the same critical temperature $T_c$, we
have either {\it both} bands normal or {\it both} bands
superconducting. In the first case, Eq. (\ref{gleqdim}b) allows only
$\Delta_j^{(1)}=0$ as a solution and thus superconductivity cannot
be restored by corrections in the extended GL model. In the
following we will consider that both bands are superconducting and
their bulk amplitudes $W_j = \sqrt{-\alpha / \beta_j}$ are real. We
rescale the equations using $\Delta_j^{(k)} = W_j
\tilde{\Delta}_j^{(k)}$ and $x = \xi \tilde {x}$, where $\xi =
\sqrt{-K/\alpha}$ is the length unit common for both condensates,
and obtain (tildes omitted)
\begin{subequations}
\begin{equation}
\Delta_j^{(0)} - [ \Delta_j^{(0)} ]^3 + \nabla^2
\Delta_j^{(0)} = 0,
\end{equation}
\begin{equation}
 \Delta_j^{(1)}  ( 1 - 3  [\Delta_j^{(0)} ]^2 ) + \nabla^2
 \Delta_j^{(1)} = \frac{F( \Delta_j^{(0)})+F_j(
\Delta_j^{(0)})}{\alpha W_j}.
\end{equation}
\label{gleq}
\end{subequations}
The first equation is the same for
both bands, while the second one differs through the terms on the
right side. This is exactly the cause for the emergence of two
different characteristic lengths in the two bands. From
Eq.~(\ref{gleq}) it is also directly apparent why this does not
happen in the order parameters of order $\tau^{1/2}$, i.e. in
Eq.~(\ref{gleq}a), but only in the higher order considerations as in
Eq.~(\ref{gleq}b). To evaluate the coefficients entering
Eq.~\eqref{gleq} one needs in principle to specify the coupling
constants $\lambda_{11}$, $\lambda_{22}$ and $\lambda_{12}$, the
partial density of states in one band, e.g. $n_1$ (since $n_1+n_2
=1$ this determines $n_2$ as well), and the ratio of the Fermi
velocities $v_1/v_2$. All other parameters enter units of scaling,
and therefore have no impact on any physical effects.
\begin{figure}[b]
\includegraphics[width=0.9\linewidth]{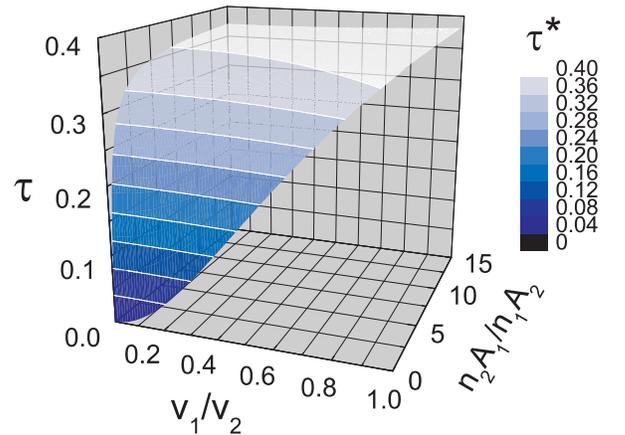}
\caption{The estimated temperature range $\tau<\tau^*$ of the
validity of the GL theory based on Eq.~\eqref{GLtauB}. The gradient
expansion made in the derivation of the GL equations (both standard
and extended) is fully justified below the plotted 3D-surface.
\label{fig1}}
\end{figure}

\paragraph*{The Ginzburg-Landau domain} Before discussing the numerical
results of Eqs.~(\ref{gleq}), it is very important to get a feeling
about the relevant values of $\tau$, i.e. {\it the domain of
applicability} of the extended GL model. First of all, our analysis
shows that $\tau \Delta^{(0)}_j \sim \Delta^{(1)}_j$. Then, by
construction, $\Delta^{(0)}_j > \Delta^{(1)}_j$ and the trivial
inequality $\tau < 1$ holds. However, this is a necessary but not
sufficient condition justifying Eqs.~(\ref{gleq}), as one also needs
to keep in mind the justification for the use of the gradient
expansion in the derivation of Eqs.~(\ref{gleq}). This expansion requires the GL coherence length $\xi$ to be larger
than the band-dependent correlation length $\zeta_j=\hbar v_j/(2\pi
T)$ that controls the spatial variations of the relevant kernels in
the integral expansion of the anomalous (Gor'kov) Green function in
powers of the band-dependent order parameters (in the clean limit),
i.e.,
\begin{align}
\zeta_1, \zeta_2 < \xi.
\label{GLtauA}
\end{align}
It is important to note here that for a two-band superconductor
$\zeta_j$ is not necessarily close to the band-dependent
generalization of the Pippard length, i.e., $\hbar
v_j/(\pi\Delta_j)|_{T=0}$, often used as an estimate of the
coherence lengths of two bands\cite{Kogan,Mosh}. Using definitions
of $\alpha$, $K$ and $\beta_j$, we obtain from Eq.~(\ref{GLtauA})
the following estimate for the GL domain $\tau \lesssim \tau^*$ with
\begin{equation}
\frac{\tau^*}{(1-\tau^*)^2} = 1+\frac{v^2_1/v^2_2-1}{1+ n_2{\cal
A}_1/(n_1{\cal A}_2)}\quad (v_1 < v_2). \label{GLtauB}
\end{equation}
If $v_1 \geq v_2$, the replacement $1 \leftrightarrow 2$ should be
made.  The Eq.~\eqref{GLtauB} shows that the extent of the
Ginzburg-Landau domain for specific multi-gap superconductor depends
on its material parameters, as was first pointed out in
Ref.~\onlinecite{Koshelev2004}. The Eq.~(\ref{GLtauB}) is most
sensitive to the ratio of the Fermi velocities $v_1/v_2$. In
particular, when $v_1/v_2=1$, the above inequality simply implies
$\tau \lesssim 0.38$, i.e., the same as in the one-band case,
regardless of the particular values of the other parameters, i.e.,
$n_2/n_1$ and ${\cal A}_2/{\cal A}_1$ (where ${\cal A}_2/{\cal A}_1=
[\Delta^{(0)}_1/\Delta^{(0)}_2]^2$). For $n_1 \approx n_2$ and
${\cal A}_1 \approx {\cal A}_2$, we typically obtain $0.27< \tau^*<
0.38$, thus the temperature domain of GL theory is still very large.
The GL domain shrinks significantly only when the ratio $n_2{\cal
A}_1/(n_1{\cal A}_2)$ acquires either extremely large or very small
values, which, e.g., can occur when the two energy gaps differ more
than an order of magnitude and at the same time $v_1$ is very
different from $v_2$. For example, for the magnesium diboride
parameters (as elaborated further below) one finds from
Eq.~(\ref{GLtauB}) $\tau ^*=0.08-0.32$ depending on the relevant
ratio of the Fermi velocities (i.e. on the direction considered,
because $\sigma$ band of magnesium diboride is highly anisotropic),
which is an order of magnitude larger than the estimate
$\tau^*\approx 0.02$ given in Fig.~1 of
Ref.~\onlinecite{Koshelev2004}. The threshold temperature for the
applicability of the GL theory ($\tau^*$) for a broader range of
parameters (i.e. other materials) is plotted in Fig. \ref{fig1}.

\begin{figure}[t]
\includegraphics[width=\linewidth]{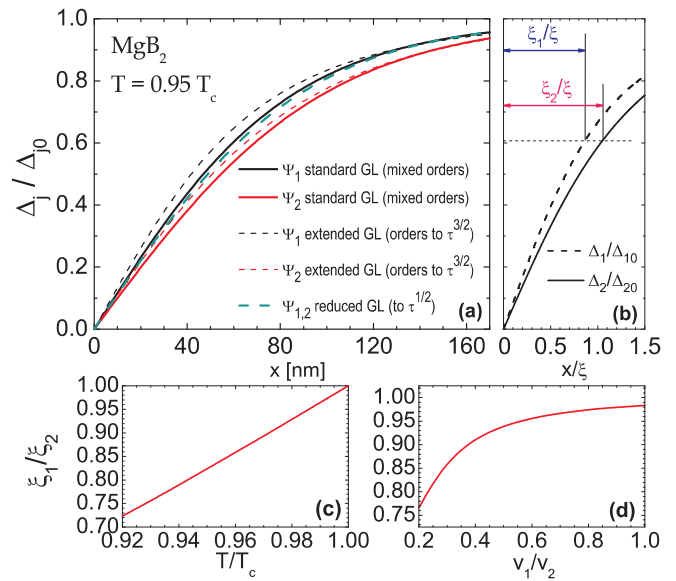}
\caption{(a) The spatial profile of the order parameters at the S/N
interface at $T=0.95 T_c$, for microscopic parameters of MgB$_2$
 ($\xi_1/\xi_2 = v_1/v_2 = 0.255$ \cite{Mosh}; $\lambda_{11}=1.88,
\lambda_{22}=0.5, \lambda_{12}=0.21$, $n_1=0.43$ \cite{Golubov}),
compared in three versions of the two-band GL theory. The ratio of
the healing lengths of the two coupled condensates $\xi_1/\xi_2$
calculated in extended GL theory [with $\xi_{1,2}$ determined as
illustrated in (b)] is shown in (c) as a function of temperature for
fixed $v_1/v_2 = 0.255$, and in (d) as a function of the ratio
$v_1/v_2$ at $T=0.95 T_c$. \label{fig2}}
\end{figure}

\paragraph*{Different length scales in selected borides, pnictides, and nanothin Pb films}
To evaluate and compare the spatial distribution of the two order
parameters in a two-band superconductor, we study the simple case of
a superconductor/normal metal (S/N) interface in the absence of any
applied magnetic field. It then suffices to consider the
one-dimensional version of Eqs. \eqref{gleq}, along the $x$-axis
perpendicular to the S/N interface. The appropriate boundary
conditions are $\Delta_j^{(k=0,1)} ( x = 0) = 0$, and $\nabla
\Delta_j^{(k)} ( x \rightarrow \infty) = 0$, where thus $x=0$ lies
at the interface, with superconductivity fully suppressed there. The
second boundary condition ensures unperturbed two-gap
superconductivity away from the S/N interface. In Fig. \ref{fig2}(a)
we show the results for $\Delta_j(x)$ in MgB$_2$ (normalized to bulk
values, at temperature $T=0.95T_c$, and with other parameters taken
from Refs. \onlinecite{Mosh} and \onlinecite{Golubov}), obtained
within standard\cite{roel}, extended\cite{Arkady}, and
reduced\cite{Kogan} GL formalism. In the case of MgB$_2$, the extended GL model clearly gives two length scales for two order parameters $\Delta_{j=1,2}$, {\it smaller} than those obtained in the standard GL theory (with
incomplete higher-order terms).

For correct comparison one must define a measure for the spatial
variations of $\Delta_j$. From the numerical solution of Eq.
(\ref{gleq}a) we found that at the characteristic distance $\xi$ the
single-gap-like order parameter $\Delta_j^{(0)}$ increases from zero
to 0.6089 of its bulk value $\Delta_{j0}$. Therefore we define the
{\it healing lengths} $\xi_j$ for the two condensates in a two-gap
superconductor using the criterion $\Delta_j(\xi_j) \equiv
0.6089\Delta_{j0}$, as depicted in Fig. \ref{fig2}(b). The
difference between the characteristic length scales in two bands is
clear already from their definition in Fig. \ref{fig2}(b), but we
emphasize this point in Figs. \ref{fig2}(c,d), where the ratio of
the healing lengths of the two order parameters is plotted as a
function of temperature [Fig. \ref{fig2}(c)] and ratio of Fermi
velocities $v_1/v_2$ [Fig. \ref{fig2}(d)]. At the lowest temperature
shown in (c), $T=0.92T_c$, the disparity between the healing lengths
is already over 25\%, and the difference increases as the ratio of
the Fermi velocities is taken smaller [see (d)].

In the analysis of the above phenomenon we noticed that only the
coefficient $S$ in front of the term $\nabla^2 \Delta_j^{(0)} $ in
$F( \Delta_j^{(0)})$ depends both on $\tau$ and $v_1/v_2$:
\begin{equation}
S = \frac{\hbar^2 N(0) W_3^2 \tau}{6\lambda_{12}}\sum_{j=1,2} v_j^2
(2 n_j \lambda_{jj}- \eta n_1 n_2 (1+2{\cal A})).
\end{equation}
Due to this special form, for particular values of $n_j$ and
$\lambda_{ij}$ the term in brackets can be positive for one band,
but negative for the other. In such a case it is possible to change
the sign of $S$ term by varying the ratio of the Fermi velocities
$v_1/v_2$ around its threshold value
\begin{equation}
\left(\frac{v_1}{v_2}\right)^* = \left ( -
 \frac{2 n_2 \lambda_{22}- \eta n_1 n_2 (1+2{\cal A})}{2 n_1 \lambda_{11}- \eta n_1 n_2 (1+2{\cal A})}
\right )^{1/2},
\end{equation}
which equals $0.46$ for the parameters of Fig.~\ref{fig2}.
Therefore, two cases with $v_1/v_2$ significantly larger and smaller
than $0.46$ will show very different behavior with respect to the
disparity of the healing lengths of the two condensates, as is
visible in Fig. \ref{fig2}(d). For $v_1/v_2\approx 0.46$, the term
with coefficient $S$ has no influence and other terms determine the
spatial behavior of the order parameters.

\begin{figure}[t]
\includegraphics[width=\linewidth]{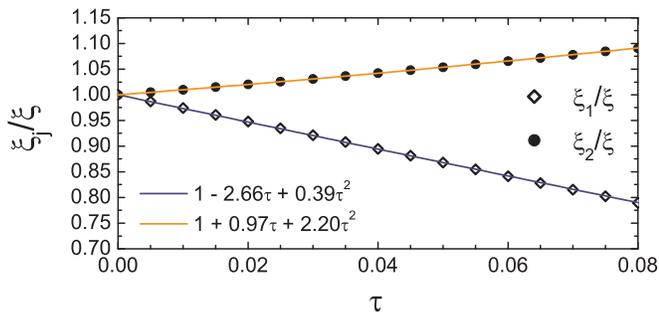}
\caption{The healing lengths of the two order parameters in MgB$_2$
[Fig. \ref{fig2}(a)] scaled to the standard GL length scale $\xi$ as
a function of temperature. The lines show the fitting functions of
the data. \label{fig3}}
\end{figure}

Besides the difference in the healing lengths of the two
condensates, the extended GL model provides insight also in a more
accurate temperature dependence of the order parameters. Namely,
standard GL theory leaves one with simply $\Delta\propto
\tau^{1/2}$, which is shown in Fig. \ref{fig3} to be clearly
inadequate for our calculations. Specifically, in Fig. \ref{fig3} we
show two healing lengths as a function of temperature, both already
normalized to the temperature-dependent length $\xi\propto
\tau^{-1/2}$. It is directly obvious that we are left with a
non-constant value, and we fitted the general temperature-dependence
of the residue to a quadratic form $\xi_j/\xi=1+A\tau+B\tau^2$, as
shown in Fig. \ref{fig3}.\footnote{Accurate fitting of our data
requires a quadratic function, although the terms of order ${\cal
O}(\tau^2)$ are not included in the extended GL theory. Therefore
the obtained coefficient $B$ should be taken with reservations.}

\begin{figure}[t]
\includegraphics[width=\linewidth]{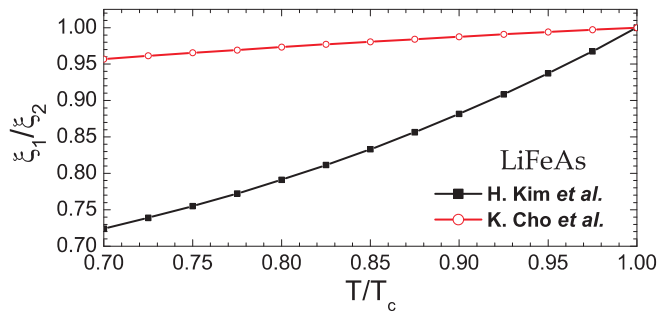}
\caption{The temperature dependence of the ratio of the healing
lengths $\xi_1/\xi_2$ in LiFeAs, for parameters from Ref.
\onlinecite{Kim} and Ref. \onlinecite{LiFeAsHc2}. \label{fig4}}
\end{figure}
In what follows, we extend our analysis to materials other than
MgB$_2$. We first address the recently discovered iron-pnictides,
more specifically LiFeAs. Interestingly enough, several existing
experimental works on this material provide us with {\it radically
different} microscopic parameters obtained from the fit of the superfluid density data in the self-consistent $\gamma$-model
\cite{gamma,Kim} and from two-band fitting of the upper critical field
\cite{LiFeAsHc2,Gurevich}. In Fig. \ref{fig4}, we show the
GL-calculated ratio of the healing lengths of two bands in LiFeAs,
for the parameters taken from latter two references. Although
different, both curves in Fig. \ref{fig4} clearly show a
discrepancy between the healing lengths, up to 20\% at temperatures
above 0.8$T_c$.

The present study and its conclusions are relevant not only for bulk
multi-band materials, but also for nanoscale superconductors, e.g.
single-crystalline nanofilms~\cite{guo}, where the multi-band
structure appears due to quantum confinement. In particular, the
${\rm Pb}(111)$ nanofilms with thickness 4 and 5 monolayers are
two-band superconductors due to the presence of only two
perpendicular single-electron levels below the Fermi energy. Our
preliminary analysis of such systems shows that the difference of
two healing lengths therein is below 5\% in the GL domain. However,
in quantum confined systems, either nanofilms or pancake fermionic
condensates, the exact position of the bands with respect to the
Fermi level (and therefore the $v_1/v_2$ ratio) can be {\it tuned}
by confinement\cite{Arkadynano} and with that the discrepancy
between the corresponding healing lengths can be significantly
enlarged.

\paragraph*{Conclusions} To summarize, we demonstrated on examples of
several multi-gap superconducting materials that characteristic
length scales of coupled condensates in two- or multi-band samples
can be significantly different from each other in the domain of the
extended Ginzburg-Landau theory. This makes our model an excellent
tool for further studies of two-band systems, where one expects a
plethora of novel physical effects emerging from the competition of
different length scales.

This work was supported by the Flemish Science Foundation (FWO-Vl),
the Belgian Science Policy (IAP), and the ESF-INSTANS network.

\end{document}